
\def\apny{{\it Ann.\ Phys.\ (New York)\ }}

\def\prd{{\it Phys.\ Rev.\ }{\bf D}}

\def\npb{{\it Nucl.\ Phys.\ }{\bf B}}

\def\prl{{\it Phys.\ Rev.\ Lett.\ }}
\def\plb{{\it Phys.\ Lett.\ }}

\def\ijmpa{{\it Int.\ J.\ Mod.\ Phys.\ }{\bf A}}
\def\mpla{{\it Mod.\ Phys.\ Lett.\ }{\bf A}}

%
%
%
\def\and{{\it\&}}
\def\half{{1\over2}}

\def\rii{\sqrt{2}}
\def\to{\rightarrow}
\def\gesim{\,{\raise-3pt\hbox{$\sim$}}\!\!\!\!\!{\raise2pt\hbox{$>$}}\,}
\def\lesim{\,{\raise-3pt\hbox{$\sim$}}\!\!\!\!\!{\raise2pt\hbox{$<$}}\,}
\def\boldoverdot{\,{\raise6pt\hbox{\bf.}\!\!\!\!\>}}

\def\ibid{{\it ibid.}\ }

\def\lcal{{\cal L}}

\def\ocal{{\cal O}}

\def\diag{\hbox{\diag}}

\def\gev{\hbox{GeV}}
\def\tev{\hbox{TeV}}

%
%
%

%
\def\inbox#1{\vbox{\hrule\hbox{\vrule\kern5pt
     \vbox{\kern5pt#1\kern5pt}\kern5pt\vrule}\hrule}}
\def\sqr#1#2{{\vcenter{\hrule height.#2pt
      \hbox{\vrule width.#2pt height#1pt \kern#1pt
         \vrule width.#2pt}
      \hrule height.#2pt}}}
\def\square{\mathchoice\sqr56\sqr56\sqr{2.1}3\sqr{1.5}3}
\def\today{\ifcase\month\or
  January\or February\or March\or April\or May\or June\or
  July\or August\or September\or October\or November\or December\fi
  \space\number\day, \number\year}
\def\pmb#1{\setbox0=\hbox{#1}%
  \kern-.025em\copy0\kern-\wd0
  \kern.05em\copy0\kern-\wd0
  \kern-.025em\raise.0433em\box0 }
\def\antes{}
\def\despues{.}
%

%
\def\sumprime_#1{\setbox0=\hbox{$\scriptstyle{#1}$}
  \setbox2=\hbox{$\displaystyle{\sum}$}
  \setbox4=\hbox{${}'\mathsurround=0pt$}
  \dimen0=.5\wd0 \advance\dimen0 by-.5\wd2
  \ifdim\dimen0>0pt
  \ifdim\dimen0>\wd4 \kern\wd4 \else\kern\dimen0\fi\fi
\mathop{{\sum}'}_{\kern-\wd4 #1}}
\def\sumbiprime_#1{\setbox0=\hbox{$\scriptstyle{#1}$}
  \setbox2=\hbox{$\displaystyle{\sum}$}
  \setbox4=\hbox{${}'\mathsurround=0pt$}
  \dimen0=.5\wd0 \advance\dimen0 by-.5\wd2
  \ifdim\dimen0>0pt
  \ifdim\dimen0>\wd4 \kern\wd4 \else\kern\dimen0\fi\fi
\mathop{{\sum}''}_{\kern-\wd4 #1}}
\def\sumtriprime_#1{\setbox0=\hbox{$\scriptstyle{#1}$}
  \setbox2=\hbox{$\displaystyle{\sum}$}
  \setbox4=\hbox{${}'\mathsurround=0pt$}
  \dimen0=.5\wd0 \advance\dimen0 by-.5\wd2
  \ifdim\dimen0>0pt
  \ifdim\dimen0>\wd4 \kern\wd4 \else\kern\dimen0\fi\fi
\mathop{{\sum}'''}_{\kern-\wd4 #1}}
%
%
\newcount\chapnum

\def\chap#1{\clearsect\clearprob
\global\advance\chapnum by 1 \par\vskip .5 in\par%
\centerline{{\bigboldiii\antes\the\chapnum\despues\ #1}}}
\newcount\sectnum
\def\clearsect{\sectnum=0}
\def\sect#1{\clearprob\global\advance\sectnum by 1 \par\vskip .25 in\par%
\noindent{\bigboldii\the\chapnum.\the\sectnum:\ #1}\nobreak}
\newcount\yesnonum

\def\verify{\global\advance\yesnonum by 1{\bigboldi (VERIFY!!)}}
\def\tocheck{\par\vskip 1 in{\bigboldv TO VERIFY: \the\yesnonum\ ITEMS.}}
\newcount\notenum

\def\note#1{\global\advance\notenum by 1{ \bf $<<$ #1 $>>$ } }
\def\noteout{\par\vskip 1 in{\bigboldiv NOTES: \the\notenum.}}
\newcount\borrownum

\def\borrow{\global\advance\borrownum by 1{\bigboldi BORROWED BY:\ }}
\def\borrowed{\par\vskip 0.5 in{\bigboldii BOOKS OUT:\ \the\borrownum.}}
\newcount\refnum

\def\ref#1{\global\advance\refnum by 1\item{\the\refnum.\ }#1}
\def\stariref#1{\global\advance\refnum by 1\item{%
               {\bigboldiv *}\the\refnum.\ }#1}
\def\stariiref#1{\global\advance\refnum by 1\item{%
               {\bigboldiv **}\the\refnum.\ }#1}
\def\stariiiref#1{\global\advance\refnum by 1\item{
               {\bigboldiv ***}\the\refnum.\ }#1}
\newcount\probnum
\def\clearprob{\probnum=0}
\def\prob{\global\advance\probnum by 1 {\medskip $\triangleright$\
\undertext{{\sl Problem}}\ \the\chapnum.\the\sectnum.\the\probnum.\ }}
\newcount\probchapnum

\def\probchap{\global\advance\probchapnum by 1 {\medskip $\triangleright$\
\undertext{{\sl Problem}}\ \the\chapnum.\the\probchapnum.\ }}
\def\undertext#1{$\underline{\smash{\hbox{#1}}}$}
\input phyzzm
\endpage
\def\gev{$GeV$}
\def\tev{$TeV$}
\def\leff{\lcal_{eff}}

\def\sutwoone{SU_2^L\otimes U_1^Y}
\def\slash{\hskip-.3cm/}
\theory
\def\today{March 23, 1993}
\pubtype{hep-ph/9303323}  
\pubnum{12} 

\titlepage

\vskip1.5in

\title{\caps{Beyond the Standard Model with Effective
Lagrangians}}\footnote{*}{Invited talk at the ``Conference on Unified Symmetry
in the Small and in the Large," Coral Gables, Jan.\ 25--27, 1993;
to be published in the Proceedings.}
\vskip.75in
\author{Martin B. Einhorn}
\address{Randall Laboratory, University of Michigan, Ann Arbor, MI 48109-1120}
\medskip
\abstract

An Effective Lagrangian description is useful for describing potential physics
beyond the Standard Model.  The method is illustrated by reference to
interactions among the electroweak bosons ($W^\pm$ \& $Z^0$).  The resulting
estimates of the magnitude of these corrections suggest that they would be at
best marginally detectable at high energy $e^-e^+$ colliders or hadron
colliders. In loop calculations, contrary to lore, such deviations from
Standard Model self-couplings never give observable corrections that grow as a
power of the scale $\Lambda$ at which new physics enters, so there are never
effects proportional to powers of $\Lambda/M_W$.

\endpage                                                           %
\singlespace

\baselineskip=15pt

\centerline{\bf{1. INTRODUCTION}}

We hope that new hadron colliders  such as LHC and SSC or future high-energy
linear $e^-e^+$ colliders (generically referred to as the NLC) will cross a
threshold for producing particles not included in the Standard Model (SM.)
Often, the hope is also expressed that the existence and nature of physics
beyond the SM might be inferred from deviations from the SM couplings, and it
is this topic of how virtual phenomena may influence observables that I shall
focus on in this talk.  I shall discuss  how one may address ``The Discovery
Potential of Future Colliders" theoretically without prejudice as to the
particular kind of new physics to be discovered.

In this regard, the interactions among the weak vector bosons are of special
interest, since nearly all theorists feel that the present theory is but an
approximation to a more fundamental description of the origin of electroweak
symmetry breaking.  One might well think that the {\it composite} nature of
longitudinal helicity states of the vectors would be revealed by non-canonical
interactions, revealing their origin in a Higgs field or a more complicated
underlying mechanism.  Thus, the study of $WW$ scattering is a primary goal
for the SSC, and $e^-e^+\to W^-W^+$ is, in any case, the largest
electron-positron annihilation channel.  In this talk, I will focus on
deviations of the triple-vector boson vertices and two-body scattering of
vector bosons.

Quite generally, such new effects can be represented as new effective vertices
involving the particles we know, but unfortunately, the language we use
depends on whether the underlying physics is decoupling or nondecoupling.  In
the former case, these are represented simply by higher dimensional operators,
while in the latter case, one must resort to an expansion in powers of
momenta, in the manner familiar from chiral perturbation
theory.\REF\georgibook{H. Georgi, {\sl Weak Interactions
and Modern Particle Theory}, Benjamin/Cum\-mings Publishing Co., Menlo Park,
CA, (1984).}[\georgibook]  Such a description has the advantage of
being model independent and process independent.  Indeed, a useful way to
compare the phenomenological implications of different models is to compare
the values implied for the parameters of the effective Lagrangian.  This
method is advantageous compared, for example, to parameterizing a ``form
factor'' associated with a particular process.  The effective Lagrangian can
be taken to respect the gauge invariance of the Standard Model.  Its
disadvantages are that it involves many parameters, since it represents all
possible models, and kinematic thresholds are not reflected, so that the
expansion will break down as one nears the threshold for new particle
production.  Such a formalism has found applications to composite fermions,
$W\,W$ scattering, especially interactions involving longitudinal $W$'s and,
it can be applied to hadrons or to quarks and leptons.  It is in fact so
general that one may wonder whether this is, in fact, a useful approach.  I
will begin by discussing the triple-vector boson interactions and later
discuss $WW$ scattering.
\medskip
\centerline{\bf{2. TRIPLE VECTOR BOSON VERTICES}}

To describe the possibilities, it has become conventional in recent years to
write down the most general Lorentz invariant Lagrangian that describes the
interaction of a photon or $Z^0$ with the $W^\pm$: For pedagogical simplicity,
we will restrict our attention to the CP-invariant terms:\REF\cpviol{F.\
Boudjema et.al.,
\prd{\bf 43} (1991) 3683; G. Gounaris, D. Schildknecht, and F.M. Renard,
\plb{\bf 263B} (1991) 291; A. DeR\'ujula et.al., \npb{\bf355} (1991) 549;  A.
DeR\'ujula et.al., \npb{\bf357} (1991) 311.}
\foot{For discussion of the phenomenology of CP-violating terms, see, e.g.,
Ref.~\cpviol.} $$\eqalign{ \lcal_{WWV}/g_{WWV}=
&\>ig_1^V (W_{[\mu\nu]}^\dagger W^\mu V^\nu-h.c.) +i\kappa_V W_\mu^\dagger
W_\nu V^{\mu\nu}\cr &+i{{\lambda_V}\over{M_W^2}} W_{[\lambda\mu]}^\dagger
W^{[\mu\nu]} V_\nu^\lambda -g_4^V W_\mu^\dagger W_\nu(\partial^\mu
V^\nu+\partial^\nu V^\mu),\cr }\eqn\peccei$$ where $W_\mu$ is the $W^-$ field,
$V_\mu$ represents either the photon $V=A$ or the $Z^0$-boson $V=Z,$
$V_{\mu\nu}=\partial_\mu V_\nu-\partial_\nu V_\mu$ and
$W_{[\mu\nu]}=\partial_\mu W_\nu-\partial_\nu W_\mu$ are the ``Abelian field
strengths."  Electromagnetic gauge invariance, generally assumed, requires
$g_{WWA}=e,$ $g_1^A\equiv1,$ $g_4^A=0.$  With the choice
$g_{WWZ}=\cot\theta_w,$ the Standard Model values are
$g_1^Z=\kappa_A=\kappa_Z=1,$ $\lambda_A=\lambda_Z=g_4^Z=0=0.$ The sensitivity
of various facilities are often quoted in terms of the limits they can place
on quantities such as $\kappa_V-1$ and $\lambda_V.$  For example, LEP-2 is
expected to restrict $\kappa_A-1$ and $\lambda$ to less than about
10\%.\REF\lepyellow{G. Barbellini et.al. in {\sl Physics at LEP,} (eds.\ J.
Ellis and R. Peccei) (CERN 86-02) vol.~2 (1986) and references
therein.}[\lepyellow] Similarly, NLC studies\REF\burke{D.L. Burke in J.
Hawthorne, ed., {\sl Gauge Bosons and Heavy Quarks,} Proceedings of the 18th
SLAC Summer Institute on Particle Physics, SLAC-REPORT-378, Stanford, 1991.}
\REF\barklow{T.L. Barklow, SLAC-PUB-5808, April, 1992 (T/E).}[\burke,
\barklow]\ estimate a sensitivity to deviations on the order of a few per cent
at 500~\gev\ and perhaps a few tenths of a percent at 1~\tev, depending on
assumptions about achievable luminosities.

This raises the question of what is the order of magnitude to be expected
for the deviations of these couplings from their SM values, the
phenomenological implications of those estimates, and the consistent use of
non-SM vertices in loop calculations.  I will argue that
\item\ (a)~the deviations of anomalous triple-vector-boson couplings from the
SM are necessarily small, since they must be associated with loop corrections
in the underlying theory,
\item\ (b)~divergences of loop integrals are either fictitious or not
observable and, hence, do not constrain these deviations.\REF\talks{J. Wudka,
talk at the {\sl Topical Conference on Precise Electroweak Measurements,}
Santa Barbara, CA, Feb.~21-25, 1991, unpublished;  M.B. Einhorn and J. Wudka,
in {\sl Proceedings of the Workshop on
Electroweak Symmetry Breaking,} Hiroshima, Nov.~12--15, 1991. (eds.\ W. Bardeen
et.al.,) Singapore: World Scientific, 1992;
M.B. Einhorn and J. Wudka, talk at ``Yale Workshop on Future Colliders,"
UM-TH-92-25 (Nov., 1992) }
\foot{I have discussed these points previously, especially at the Yale
Workshop last October.[\talks]}
\medskip
\centerline{\bf{3. EFFECTIVE LAGRANGIANS}}

Quite generally, effective Lagrangians are non-renormalizable by virtue of
higher-dimensional interactions, such as the $\lambda_V$ term, but ${\cal
L}_{WWV}$ is also non-renormalizable because it introduces dimension-four
vertices, such as $\kappa_V,$ in a manner that explicitly breaks the
$\sutwoone$ electroweak gauge invariance.  Indeed, it is regarded as an
experimental goal to establish or disprove that these couplings are equal to
those of the SM.  So long as one is not imposing $\sutwoone$ gauge invariance,
one ought to include in the effective Lagrangian explicit mass terms for the
vector fields, $$M_W^2 W_\mu^\dagger W_\mu+{{M_Z^2}\over2} Z_\mu
Z_\nu.\eqn\massterms$$  Thus, the first question for such a general approach
should be why the SM provides such a good prediction for the relation between
these masses (given the Fermi constant $G_F$ and the fine structure constant
$\alpha.$)  The usual answer, at least in technicolor models, is that the
Higgs sector respects an approximate global, weak isospin symmetry, called
custodial $SU_2$,\REF\suss{Sikivie, Susskind, Voloshin, Zacharov} broken only
by hypercharge and quark masses, but, especially in the aftermath of the LEP-1
results, we are confident of the SM at accuracies at the level of SM one-loop
corrections, so a discussion of custodial symmetry is somewhat beside the
point of this talk.

But the general question raised is how can one understand the success of the
SM, including the one-loop corrections important for interpreting LEP
experiments, if one does not have $\sutwoone$ gauge invariance.\REF\dghm{A.
De~R\'ujula, M.B. Gavela, P. Hernandez, and E.  Mass\'o, CERN-TH-6272-91 (Oct.
1991.)}\foot{This has been discussed at some length in Ref.~\dghm.}  This is
not quite the right or meaningful question, since one may, without loss of
generality, assume that the effective Lagrangian, Eqs.~\peccei\ and
\massterms, is an expression of a gauge invariant effective Lagrangian in a
unitary gauge.\REF\blone{C.P. Burgess and D. London, McGill-92-04 (Mar.\
1992.)}\foot{This point has been made previously in Ref.~\talks\ and
emphasized in Ref.~\blone.}  This observation, while correct, is somewhat off
the mark, since one may regard a Lagrangian that appears to be
non-gauge-invariant as the unitary gauge expression of a gauge-invariant
Lagrangian corresponding to gauge symmetries other than $\sutwoone.$ The
electroweak $\sutwoone$ is not singled out by this observation, so we must
probe more deeply to discover the significance of the SM gauge symmetry.

Unfortunately, the discussion necessarily bifurcates, depending on the nature
of the underlying physics.\foot{Often times, these are referred to as the
strongly interacting, nondecoupling scenario  and the weakly interacting,
decoupling scenario.}  In general, the most conservative approach, which can
always be used for energies small compared to the scale of new physics,
is to expand Green's functions in powers of momenta, so-called
chiral perturbation theory.\REF\chiralpert{M. Golden and L. Randall,
\npb{\bf 361} (1991) 3;  B. Holdom and J. Terning, \plb{\bf247B} (1990) 88; B.
Holdom, \plb{\bf259B} (1991) 329.} \REF\drv{J.F. Donoghue, C. Ramirez, and G.
Valencia, \prd{\bf 39} (1989) 1947; J.F. Donoghue and C. Ramirez, \plb{\bf
234B} (1990) 361.}
\REF\holdomee{B. Holdom, \plb{\bf258B} (1991) 156.}
\REF\bdv{J. Bagger, S. Dawson, and G. Valencia, \prl{\bf67} (1991) 2256 and
FERMI\-LAB-PUB-92/75-T (Aug., 1992, revised).}[\chiralpert, \holdomee, \bdv.]

The prototypical underlying model is technicolor, in which the natural scale
of the expansion is {\it at most} $4\pi v,$ where $v\approx 250~GeV$ is the
weak scale.  In such a case, the new physics is strongly interacting, and its
scale is set by the ratio of this strong interaction to the gauge
interactions.  Inevitably, it seems, there are technifermions and/or
scalar pseudo-Goldstone bosons in the hundreds of GeV mass range, so this is
not to say that thresholds for new physics are necessarily in the multi-TeV
range.  There could also be new physics at higher scales, such as the extended
technicolor interactions, or a fourth fermion generation of fermions (which,
because of the invisible $Z^0$ width, would necessarily involve a heavy
neutrino.)

If, on the other hand, there is a relatively light Higgs boson ($m_H\lsim
800~GeV,$) then a more appropriate description of electroweak
symmetry-breaking is, as in the SM, with an explicit Higgs field treated
perturbatively.  Although there may be other relatively light particles, the
intrinsic scale of new physics may not be directly associated with electroweak
symmetry breaking and may even be higher than the weak scale, such as the
scale of supersymmetry(SUSY) breaking in ``low-energy" SUSY models.

Finally, of course, both kinds of new physics may be present, as in
technicolor models with extended technicolor interactions at much higher
scales responsible for fermion masses.

In any case, the SM may be understood as the leading behavior in a
systematic expansion of a more comprehensive theory applicable at higher
energy scales.  Familiar examples are the manner in which QED emerges from a
unified electroweak theory below 80~\gev\/ or the way chiral perturbation
theory for pions emerges from QCD below 1~\gev.  As a result, contrary to what
you might think, the behavior of vector boson interactions below the weak scale
will not be wildly different in theories in which $W$ bosons are in some sense
composite. Light vector bosons cannot behave other than gauge
bosons.\REF\veltman{M. Veltman, Acta Phys.\ Polon.\ B12 (1981) 437.}\foot{This
point of view has been elaborated more than ten years ago by Veltman[\veltman]\
and is closely related to the notion of naturalness.} The so-called delicate
gauge theory cancellations are intimately linked with the fact that the vector
masses are small compared to the scale of their compositeness.  So, in a
sense, gauge invariance is not really the relevant issue; the question is {\sl
whether we may regard the weak vector bosons as elementary gauge particles
over a range of momenta large compared to their masses.}  This is what is
tacitly assumed in performing radiative corrections in the SM, and these
calculations and their consequences will break down if the cutoff on momentum
integrals is not much greater than the vector boson masses.  This is
illustrated by technicolor models, which are elegant and reasonably
unambiguous expressions of the notion of a composite Higgs field, where the
scale of new physics is expected to be around $1~TeV,$ an order of magnitude
larger than the vector masses.

Let me first describe non-SM triple-vector-boson couplings.
Because the description in terms of an elementary Higgs is somewhat more
familiar to most people than chiral perturbation theory, I will use the linear
language, but the general conclusions that I will draw for triple-vector-boson
interactions are not qualitatively different from the nonlinear scenario, to
which I shall return in discussing $WW$ scattering.  In this case, the
corrections to the SM Lagrangian are simply gauge-invariant,
Lorentz-invariant, higher dimensional operators involving the Higgs scalar,
vector, or fermion fields.  In principle, there can be more than one Higgs
doublet, as in SUSY models, but it is not necessary to consider that
alternative for our present purposes.  The first corrections coming from
physics at higher scales that modify the vector boson properties are dimension
six operators,\REF\bw{W. Buchm\"uller and D. Wyler, \npb{\bf268} (1986) 621; a
term omitted can be found in W. Buchm\"uller, B. Lampe, and N. Vlachos,
\plb{\bf197B} (1987) 379.}[\bw] such as : $$ \eqalign{ \ocal_\phi^{(3)} &=
|\phi^\dagger D_\mu\phi|^2,\cr
\ocal_W &= \epsilon_{ I J K }{ W_\mu }^{ I \nu
} { W_\nu }^{ J \lambda }{ W_\lambda }^{ K \mu }, \cr \ocal_{ W B } &= (
\phi^\dagger \tau^I \phi ) { W_{ \mu \nu } }^I B^{ \mu \nu }, \cr \ocal_{ \phi
W } &= \half ( \phi^\dagger \phi ) W_{ \mu \nu }^I W^{ I \mu \nu }, \cr \ocal_{
\phi B } &= \half ( \phi^\dagger \phi ) B_{ \mu \nu } B^{ \mu \nu } , \cr }
\eqn\dimsix$$
where $\phi$ is the usual Higgs doublet.  These add to the SM
Lagrangian terms of the form $${1\over{\Lambda^2}}\sum
\alpha_k\ocal_k.\eqn\leffsix$$ Here, $\Lambda$ represents a generic scale of
new physics and each $\alpha_k$ may be thought of a coupling constant
associated with the corresponding operator.\foot{For simplicity, we have
restricted our attention here to CP-conserving operators, but the CP-violating
operators have also been delineated.[\bw]}  For convenience, we use the same
scale $\Lambda$ for each operator, but, of course, this is simply a matter of
convention, and the threshold for new particle production could be different
from $\Lambda.$

These operators contribute a variety of corrections to the SM.  To make
contact with the non-gauge invariant formalism of Eq.~\peccei, one may go to
the unitary gauge\foot{One can also make contact with chiral perturbation
theory by setting $\Lambda=v,$ in which case these operators are a subset of
those that arise from the chiral-dimension two and four operators.  See
Section~6 below.} $$\phi = {1\over\rii}\pmatrix{0\cr
v+\sigma\cr},\eqn\unitary.$$  Obviously, if we can replace the Higgs field by
its VEV $v$, the last two operators are simply wave function renormalizations
and have no observable consequences.  This means that, until Higgs bosons are
observed, these operators can be ignored at tree level.
Similarly, unless one observes Higgs boson interactions directly, the first
operator, $\ocal_\phi^{(3)}$, simply contributes to $M_Z^2,$ altering the
well-confirmed relationship expressed by the $\rho$ parameter.\REF\mybook{M.B.
Einhorn (ed.,) {\sl The Standard Model Higgs Boson,} Amsterdam: North-Holland,
1991.}\foot{Here $\rho\equiv M_W^2/M_Z^2\cos^2\theta_w.$  For a review of its
significance, see Chapter~1 of Ref.~[\mybook.]}   So this contribution from
new physics must be small, less than about $1\%$ of $M_Z.$\foot{Once the top
quark mass is known, the limit will drop by almost another order of
magnitude.} Other effects can be easily worked out, for example,
$$\eqalign{\kappa_A-1=& { {\alpha_{ W B }}\over{\Lambda^2} }{
{4M_W^2}\over{gg'} },\cr { {\lambda_A}\over{M_W^2} }=&{6\over g}{
{\alpha_W}\over{\Lambda^2} },\cr}
\eqn\correspond$$
with similar expressions for $g_1^Z,$ $\kappa_Z$ and $\lambda_Z.$

As a theoretical aside, there are several technical subtleties associated with
the elaboration of higher dimensional operators.  One is whether the effective
Lagrangian may be assumed to be gauge invariant, given that the effective
action is gauge dependent in general.  Fortunately, the answer is in the
affirmative.\REF\arzt{C. Arzt, UM-TH-92-28, (Mar.\ 1992).}[\arzt]

Another point has to do with the fact that not all higher dimensional
operators are physically distinct, since one may use the classical equations
of motion to relate some operators to others.  Generally, this application
of the classical equations of motion has been said to be limited to tree
approximation, but in fact, one may use them even though the resulting
effective Lagrangian is to be used in loop calculations.[\arzt]  Certain
operators, such as $\ocal_W,$ that contribute new three-point and higher
vertices may be related to others that modify the two-point function (vacuum
polarization tensor.)  A further complication is precisely how the equations
of motion are to be applied in spontaneously broken theories.  In the final
analysis, the answer is quite simple:  one may assume that the form of the
effective Lagrangian is gauge invariant and ignore spontaneous symmetry
breaking in the application of the classical equations of motion.[\arzt]  The
bottom line is that one may simply drop operators involving scalar or vector
fields on which a d'Alembertian ($\square$) acts or involving fermion fields
on which $D\slash$ acts.  Thus, one may arrive at a basis set of operators
that are independent under application of the classical equations of motion.

De~R\'ujula et.al.[\dghm]\ have emphasized that, depending on the
order in which experiments are done, some basis sets may be better than
others.  They have conjectured that it would be unnatural for new physics to
yield only those operators that contribute anomalous vertices without also
modifying the vacuum polarization tensor.  Thus, this seemingly technical
issue has physics content, inasmuch as some of the operators unrestricted by
LEP-1 measurements are not independent of operators that are restricted in the
absence of fine tuning.  For the present, one must be clear that the
conclusion of Ref.~\dghm, that LEP-1 restrictions obviate LEP-2 measurements
of triple vector boson vertices, is a direct consequence of this assumption.
We do not know whether this conjectured extension of the notion of naturality
is correct, but we have not found a model that provides a counterexample and
believe it is very likely true.  On general grounds, we shall argue below that
the deviations from the SM will be too small to be observed at LEP-2.
\medskip
\centerline{\bf{4. ORDERS OF MAGNITUDE}}

Assuming that $\Lambda$ does represent a new heavy particle mass, how large
would we expect the various couplings $\alpha_k$ to be?  There is certainly
one power of $g$ or $g'$ for each power of the corresponding field operator
$W^I$ or $B_\mu,$ but an essential question governing their magnitude is
whether we can expect contributions to these operators at tree level in some
models or whether they invariably represent loop corrections.  For the
operators discussed in the preceding section, the only tree diagram
contributions would be from heavy vectors or scalars, and one can show that
there can be no tree contributions compatible with $\sutwoone$
symmetry.\REF\aewtwo{Arzt, Einhorn, Wudka, in preparation.}[\aewtwo]
Therefore, all these operators arise from one-loop corrections at best, so
that it is natural to associate a factor of ${1/{16\pi^2}}$ should be
associated with each.  As a result, so long as $\Lambda\gsim v,$ the presence
of such terms are not really troublesome for the SM.  Consider for example,
$\ocal_\phi^{(3)}.$  This can get a contribution from essentially any one loop
correction that breaks the custodial $SU_2$ symmetry; for example, the top
quark makes an important contribution to $\alpha_{\phi^{(3)}}.$  Because this
is proportional to the square of the Yukawa coupling of the top to the Higgs,
the limits on its contribution to $M_Z$ place an upper limit of about
$180~GeV$ on $m_t.$\REF\georgitwo{H.\ Georgi, HUTP-90/A077 (March,
1991.)}\foot{For a discussion of the usual SM radiative corrections in the
language of effective field theory, see Ref.~\georgitwo.}

Our interest for collider physics has more to do with the contributions of the
two other operators $\ocal_W$ and $\ocal_{ W B }$ in Eq.~\dimsix.  The
preceding reasoning leads to the estimates $$ \alpha_W \sim { g^3 \over 16
\pi^2 } , \quad \alpha_{ W B } \sim { g g' \over 16 \pi^2 }.\eqn\estimates$$
Correspondingly, one finds from Eq.~\correspond\ that, taking $\Lambda\approx
v,$ $|\kappa-1|\sim3\times 10^{-3} $ and $ |\lambda|\sim2\times 10^{-3}$.
These estimates are an order of magnitude smaller than the sensitivity claimed
for an NLC at 500~\gev\ and are borderline for an $e^-e^+$ collider at
1000~\gev!  It should be said that, if one takes QCD as a prototype of
technicolor,\REF\gasserleut{J.\ Gasser and H.\ Leutwyler, \apny{\bf158} (1984)
142 and \npb{\bf250} (1985) 465.}[\gasserleut] it turns out that these sort of
back-of-the-envelope estimates of couplings are in agreement with some and are
larger than most of the experimental values for coefficients in the chiral
perturbation expansion.\foot{For a detailed analysis of $e^-e^+\to W^-W^+$ in
the language of chiral perturbation theory, see Ref.~\holdomee.}   In some
models, it may also be the case that there are many similar contributions at
approximately the same mass scale that add constructively, enhancing the
one-loop estimate.  But, even allowing an order of magnitude increase, these
estimates remain at the borderline of the anticipated sensitivity of a
500~\gev\ $e^-e^+$ collider.

A higher energy $e^-e^+$ collider may achieve greater sensitivity, but, at
1~\tev, if one has not crossed a threshold for new particle production, then
surely the relevant scale of new physics $\Lambda$ will also be larger and,
correspondingly, $\kappa-1$ and $\lambda$ smaller.  This illustrates one
problem with probing virtual effects by going to higher energy: your figure of
merit is a moving target.  Not that I am against going to higher energy; the
best way to study new physics is to cross its threshold and look directly at
it.  While an NLC at 500~\gev\ may be a fine facility for many things, it
seems unlikely to be very useful for probing anomalous vector boson couplings.

With regard to the sensitivity of LHC or SSC to anomalous triple vector boson
couplings, I am frankly uncertain.  A recent study in the context of chiral
perturbation theory, performed by Falk, Luke, and Simmons,\REF\fls{A.F.\ Falk,
M.\ Luke, and E.H.\ Simmons, \npb{\bf365} (1991) 523.}[\fls] suggests that the
sensitivity to ``anomalous couplings" will be no greater than at an NLC at
500~\gev, a somewhat surprising conclusion.  This results from the assumption
that, because of QCD jet backgrounds, hadronic decay modes of $W^\pm$ or $Z^0$
will not be identifiable, about which there is some dispute.\REF\gunion{J.
Gunion et.al.,
\prd40 (1989) 2223.}[\gunion]  If the hadronic decays were visible, then the
SSC sensitivity could be increased.  This is a topic requiring further
detailed study.
\medskip
\centerline{\bf{5. LOOP CORRECTIONS}}

The contributions of anomalous vector boson interactions to radiative
corrections have often been used to constrain these vertices.\REF\loops{M.\
Suzuki, \plb{\bf153B} (1985) 289; J.J. van~der~Bij,
\prd{\bf35} (1987) 1088; A. Grau and J.A. Grifols, \plb{\bf233B} (1986) 233;
J.A. Grifols, S. Peris, and J. Sol\'a, \plb{\bf197B} (1987) 437,
\ijmpa{\bf3} (1988) 225; \splitout
G.L. Kane, J. Vidal, and C.-P. Yuan, \prd{\bf 39} (1989) 2617; G.\
B\'elanger, F. Boudjema, and D. London, \prl{\bf65} (1990) 2943; R.\
Alcorta, J.A. Grifols, \mpla{\bf2} (1987) 23; B. Altarelli and R. Barbieri,
\plb{\bf253} (1991) 161; F. Boudjema et.al., \prd{43} (1991) 3683; H.
Neufeld, J.D. Stroughair, and D. Schildknecht, \plb{\bf 198B} (1987)
563.}[\loops]
The point is that, when one performs loop integrals starting from the
non-gauge invariant form, Eq.~\peccei, one often finds that they diverge as a
power of momentum.  The origin of these divergences are threefold: (1)~the
factor of $k^\mu k^\nu/M_W^2$ in the vector boson propagators, (2)~the lack of
``gauge cancellations" when vertices such as $\kappa_V$ do not take their SM
values, and (3)~the higher-dimensional vertices such as $\lambda_V$ that
contain additional powers of momenta.  As I shall explain, {\bf all claims
concerning the observable  sensitivity of radiative corrections to powers of a
cutoff are wrong!} \REF\bltwo{C.P. Burgess and D. London, McGill-92/05
(Mar.~1992,) 92/38 (Sept.~1992,) 92/39 (Sept.~1992.)}\foot{This has been
emphasized in Ref.~\talks\ and again recently in Ref.~\bltwo.} Unfortunately,
such analyses have frequently been cited as one justification for
LEP-2,[\lepyellow]\ SSC, or NLC, which might reduce limits by another
order of magnitude or more.  (See the discussions in Refs.~\burke\ and
\barklow.)

First of all, we have remarked and illustrated that the non-gauge-invariant
expression Eq.~\peccei\ may be regarded as the unitary gauge result of a
general gauge-invariant effective Lagrangian.  Therefore, those power-law
divergences that are associated with the $k^\mu k^\nu/M_W^2$ terms in the
propagators are unitary gauge artifacts, since in any renormalizable gauge,
these are replaced by terms behaving as $k^\mu k^\nu/k^2$ for large momenta.
This dispenses with (1) and (2) above.  Secondly, while vertices associated
with higher-dimensional operators will lead to power divergences, these are
compensated by corresponding powers of $1/\Lambda$ extracted in front.  So,
higher-dimensional operators do not lead to net increases in the dependence on
the cutoff.  The only quantities that remain sensitive to the scale of new
physics are scalar mass parameters; that is the usual naturalness problem.
Finally, such divergences, even logarithmic ones, are not in principle
observable; power-counting arguments similar to proofs of renormalizability
show that these divergences are the coefficients of local operators, so they
simply renormalize the couplings associated with other local operators already
present.  (The effective Lagrangian already contains all possible operators
consistent with the symmetries of the theory.)  So, as usual, such divergences
can only be used to argue about the {\it natural} size of such couplings, just
as the quadratic divergence in the scalar mass of the Higgs field restricts
the scale $\Lambda$ of new physics to be naturally less than about $4\pi
v.$\REF\thooft{G.\ 't~Hooft, in {\sl Recent Developments in Gauge
Theories,}(eds.\ G.\ 't~Hooft et.al.,) Plenum Press, NY, 1980.}[\thooft]
Concerning the logarithmic divergences, they may be interpreted as
contributing to the $\beta$ functions for other renormalized operators and so,
the way in which operators mix and coupling constants evolve from the cutoff
$\Lambda$ down may be of interest.  In practice, however, these are invariably
small effects for the range of $\Lambda$ to which experiments are actually
sensitive.\REF\aew{C. Arzt, M.B. Einhorn, and J. Wudka, UM-TH-92-17 (August,
1992.)}\foot{We have illustrated the correct use of loop corrections in the
case of past and future measurements of $g_\mu-2,$ a truly high precision
experiment.[\aew]}
\medskip
\centerline{\bf{6. V\ V-SCATTERING IN HADRON COLLIDERS}}

Now let me turn to a discussion of two-body scattering of weak-vector-bosons
in hadron colliders.  Given that triple-vector boson vertices are so difficult
to observe, one would not think that anomalous quadruple vector boson vertices
would be any easier.  The answer is in fact rather complicated and depends on
a number of factors.  Quite generally, at energies large compared to their
mass, longitudinal vector bosons retain the memory of their origin in the
Higgs mechanism and behave more like scalar particles than massless vector
particles.  This behavior is embodied in the equivalence theorem,
\REF\equivthm{J.M.\ Cornwall, D.N.\ Levin, and G.\
Tiktopoulos, \prd{\bf16} (1977) 1519; B.W.\ Lee, C.\ Quigg, and H.\ Thacker,
\prd{\bf16} (1977) 1519; M.S.\ Chanowitz and M.K.\ Gaillard, \npb{\bf261}
(1985) 379; H.\ Veltman, \prd{\bf41} (1990) 2294.}[\equivthm]
one consequence of which can be stated by comparison with the situation when
the gauge coupling is set zero:  The equivalence theorem tells us that the
interactions of longitudinal vector bosons go smoothly over to those of the
Goldstone bosons that were eaten when gauge interactions were switched on.

First of all, we should remind ourselves of the situation within the SM.  At
energies large compared to the Higgs mass $m_H,$ the scattering amplitude for
longitudinal vectors is that of four scalars, which in tree approximation is
simply their self-coupling $\lambda=m_H^2/2 v^2.$  Thus, even for
$\lambda/16\pi^2$ not large, i.e. even within the perturbative
regime,\foot{The breakdown of perturbation theory occurs for $m_H\approx
1~TeV$.} the interaction strength can be much larger than its naive value
$g^2$, characteristic of massless or massive but transverse vector bosons.
And, if there is a Higgs boson or other thresholds in the hundreds of \gev, it
is likely that the SSC or, possibly, the LHC will find evidence for the
production of new particles.\REF\hunt{J.F. Gunion et.al., {\sl The Higgs
Hunter's Guide,} Benjamin/Cummings Publishing Co., Menlo Park, CA, USA
(1990).}[\hunt]

If, on the other hand, the threshold associated with the physics of
electroweak symmetry breaking is very large (sometimes described as the heavy
Higgs limit) compared to accessible energies,\foot{It must be understood that
the maximum energy of the SSC is not accessible, either because the parton
distribution functions fall as the parton momentum increases or because the
relevant cross sections are small, or both.}  then the question becomes
whether its effects can be detected at lower energies.  In this extreme, the
amplitude is described by chiral perturbation theory, to be reviewed below.
In lowest order, the amplitude for longitudinal vector boson scattering is
proportional to $(E/v)^2$, where $E$ is the center-of-mass energy.  In fact,
this reaches the unitarity limit on the real part of an S-wave scattering
amplitude at $E\approx 1.2~TeV.$\REF\luscher{M. L\"uscher and P.
Weisz, \plb{\bf 212B}(1988) 472.}[\luscher]
This result can be obtained by direct calculation in the SM for $E\ll m_H,$
but, given that $\rho=1$, this behavior below the scale $v$ is universal, a
kind of low-energy theorem resulting from the $SU_2^L$ symmetry of the
SM.\REF\lowenergy{M. Chanowitz, M. Golden, H.  Georgi, \prl{\bf 57}, 2344
(1986,) \prd{\bf 36}, 1490 (987.)}[\lowenergy]

To sum up, it is generally believed that the SSC will either produce and
detect a Higgs boson or discover a strongly interacting Higgs
sector.\REF\nolose{M. Chanowitz, in {\sl DPF '87, Proceedings of the Salt Lake
City Meeting,} (eds.\ C. DeTar and J. Ball,) World Scientific, Singapore,
1987.}[\nolose]  Of interest to us here is the latter case in which we must
infer the properties of the Higgs sector from virtual effects.  The expression
of an appropriate effective Lagrangian requires a nonlinear representation of
the underlying $\sutwoone$.  To this end, one defines a field $\Sigma$ in
terms of three would-be Goldstone bosons $w_i$ according to $$ \Sigma\equiv
\exp{ ( { { i\vec{\tau}\cdot\vec{w} } / {v} } ) },\>\>
V_\mu\equiv\Sigma^\dagger (D_\mu\Sigma),$$ where
$D_\mu\Sigma=\partial_\mu\Sigma-i{{g_2}\over2}W_\mu\Sigma-
i{{g_1}\over2}\Sigma B_\mu\tau_3.$ Here, $\vec{\tau}$ are the Pauli matrices,
$W_\mu$ is the matrix of $SU_2$ gauge fields ($W_\mu\equiv W_\mu^a\tau^a/2$)
and $B_\mu$ is the hypercharge field.  By construction, $\Sigma$ transforms
linearly under an $\sutwoone$ transformation as $$\Sigma\to e^{ ( {
{-i\vec{\theta_L}\cdot\vec{\tau} } / {2} } ) }\,\Sigma\, e^{ ( { {
i{\theta_R}\tau_3}  / {2} } ) }.$$  The effective field theory, called chiral
perturbation theory,\REF\gassleut{J. Gasser and H.  Leutwyler, {\sl Ann.
Phys.} {\bf 158} (1984) 142; {\sl Nucl. Phys.} {\bf B250} (1985) 465; \ibid
{\bf B250} (1985) 517; \ibid {\bf B250} (1985) 539.  S.  Weinberg, {\sl
Physica} {\bf A96} (1979) 327.}[\gassleut, \georgibook] is simply a momentum
expansion that, in coordinate space, corresponds to an expansion in powers of
the covariant derivative $D_\mu.$  Note that the gauge fields are to be
counted as dimension 1, the same as the ordinary derivative, which is correct,
at least for the longitudinal degree of freedom.  The ``chiral-dimension-two"
and ``-four" terms take the following form:  $$\eqalign{
\leff =&\lcal_2  + \lcal_4\cr  \lcal_2\equiv& {{v^2}\over 4} {\rm
Tr}\bigl(V_\mu V^\mu\bigr)+ \delta\rho {{v^2}\over8}\bigl[{\rm
Tr}(\tau_3V_\mu)\bigr]^2,\cr  \lcal_4\equiv& { {L_1}\over{16\pi^2}}\,
{\bigl[{\rm Tr\,}(V_\mu V^\mu)\bigr]}^2 +  { {L_2}\over{16\pi^2}}\,
{\bigl[{\rm Tr}\,(V_\mu V_\nu)\bigr]}^2\cr &-ig{ {L_{9W}}\over{16\pi^2}
}\,{\rm Tr}\bigl[W^{\mu\nu}D_\mu\Sigma D_\nu\Sigma^\dagger\bigr]  -ig'{
{L_{9B}}\over{16\pi^2}}\,{\rm Tr}\bigl[B^{\mu\nu}D_\mu\Sigma^\dagger
D_\nu\Sigma\bigr]\cr &-igg'{ {L_{10}}\over{16\pi^2}}\,{\rm
Tr}\bigl[B^{\mu\nu}\Sigma^\dagger W_{\mu\nu} \Sigma\bigr]\cr}.\eqn\chiral$$
In these expressions, the field strength tensors are given by
$B_{\mu\nu}\equiv(\partial_\mu B_\nu-\partial_\nu B_\mu)\tau_3$ and
$W_{\mu\nu}\equiv\partial_\mu W_\nu-\partial_\nu W_\mu-ig[W_\mu,W_\nu].$ In
some ways, the physical significance of the various terms may be interpreted
most directly in the ``unitary gauge" where $\Sigma=1.$  Thus, it would appear
that the first term of $\lcal_2$, the ``kinetic energy," contributes only to
the vector masses.  However, this term contains more information than that,
since, in the limit that the gauge couplings are zero, it also contributes to
the scattering of Goldstone bosons which, by the equivalence theorem, is equal
to the scattering of longitudinal gauge bosons. Thus, this term includes the
low-energy theorem we referred to earlier, even though paradoxically, in the
unitary gauge, it appears to describe no scattering at all.  The second term
in $\lcal_2$ term with coefficient $\delta\rho$ is obviously like the term we
called $\ocal_\phi^{(3)}$ in the linear case.  Thus we know that $\delta\rho$
is experimentally restricted to be $\lsim 1\%,$ which is evidence of the
approximate $SU_R$ custodial symmetry of the strongly interacting Higgs
sector.  The terms $L_9$ and $L_{10}$ involve triple vector boson couplings,
contributing to $\kappa-1$, while $L_1$ and $L_2$ contribute to the scattering
of longitudinal vectors.  As this latter is our present interest, we will
ignore all but $L_1$ and $L_2$ for now.
We have extracted appropriate powers of $16\pi^2$ according to the usual
prescription.\REF\georgithree{H. Georgi and A. Manohar, \npb{234} (1984) 189;
H. Georgi, HUTP-92/A036 (July, 1992.)}\hfill[\georgibook, \georgithree]
However, in principle, it is not obvious that $4\pi v$ sets the scale of the
chiral perturbation expansion and that these terms as small as indicated.
However, as mentioned earlier, it does work in hadron physics, i.e., the scale
of the momentum variation is set by the vector mesons $\rho, \omega, \phi,$
and is the reason why the vector-meson-dominance model works so well.  Thus
the scale of the breakdown of the chiral perturbation expansion is on the
order of $m_\rho$, which is only about half of $4\pi f_\pi.$  Indeed, the
famous fits of Gasser and Leutwyler[\gassleut]\ confirm that the $L_i$ are of
order one, in particular, $L_1 = 0.9\pm 0.3$ and $L_2 = 1.7\pm 0.7.$  Thus,
this ``naive dimensional analysis" does work.

If one can measure $L_1$ and $L_2$, one can discriminate somewhat between
alternate underlying dynamical mechanisms.[\drv, \bdv]  For example, a heavy
scalar (Higgs) gives $L_1 = (64\pi/3) (v^4\Gamma_H/m_H^5),\>L_2 = 0,$ whereas
a heavy vector meson (techni-$\rho$) gives $ -L_1 = L_2 =
(192\pi)(v^4\Gamma_\rho/m_\rho^5),$ a very different pattern.  For example,
for the SM Higgs with $m_H=1~TeV,$ we find $L_1\approx 0.13,$ with the others
quite negligible.  For a heavier Higgs, it may be even larger, but this is a
strongly interacting regime about which we cannot speak with confidence.  As
another example, for a technirho with mass 1.5--2.0~\tev, typically $|L_1|$
and $|L_2|$ are on the order of a few tenths, whereas $L_9$ and $L_{10}$ are
of $O(1).$  In fact, the situation may be slightly better, since our order of
magnitude estimates apply at the scale of new physics $\Lambda$ and, at a
lower scale $\mu$, they may be logarithmically enhanced by a factor of
$\ln(\Lambda/\mu).$

The question is whether numbers such as these are large enough to be
detectable.  Just how well SSC experiments can determine $L_1$ and $L_2$ is
not known, because it is not now understood how efficiently weak vector bosons
can be observed in their non-leptonic decay modes.  A rough estimate of the
sensitivity[{\fls, \bdv}]\ suggests that, relying solely on leptonic decay
modes, the triple vector boson couplings can be detected only if $L_9$ or
$L_{10}$ are larger than about $3.$  This is, as we have said before, quite
inauspicious.

Similarly, $L_1$ and $L_2$ must be greater than about 1 or so to be
detectable, so these are slightly more interesting limits.  In this regard, it
seems that the $W^+W^+$ mode is particularly attractive,\REF\plusplus{M.
Berger and M.  Chanowitz, \plb{\bf 263B} 509 (1991); V Barger et.al., \prd{\bf
42} 3052 (1990); D. Dicus et.al., \plb {\bf 258B} 475 (1991)}[\plusplus, \bdv]
inasmuch as the backgrounds are less and the sensitivity to these two
parameters greatest.  The conclusion is that, while the SSC may be good enough
to determine whether the Higgs sector is strongly interacting, unless a new
threshold is crossed, it will be very difficult to ascertain the nature of the
underlying dynamics.
\medskip
\centerline{\bf{7.  CONCLUSIONS}}

I want to emphasize once again that there is content to saying that the weak
vector bosons are associated with gauge fields, and the equivalence of
Eq.~\peccei\ to the unitary gauge expression of a gauge-invariant Lagrangian
may be misinterpreted.  This equivalence does not imply that every vector
particle in the world is necessarily associated with a gauge field.    While
the effective interactions of composite particles may be written in gauge
invariant form, if the Lagrangian can only be used in tree approximation, that
is irrelevant.  The essence of a gauge particle is that is may be treated as
elementary over a momentum range large compared to its mass, so that its
composite structure, if any, is not revealed up to some scale of new physics.
The gauge symmetry is irrelevant if this new scale is comparable to the vector
masses.  Contrast the situation we find with the $W^\pm$ and $Z^0$ to weakly
bounds systems such as orthopositronium or the $J/\psi,$ whose substructure is
revealed at momenta small compared to their masses.  Similarly, I would argue
that the $\rho$ meson is not a gauge particle.  For a gauge particle,
loop calculations may be reliably performed up to the scale at which its
substructure must be taken into account.  People who wish to entertain that
the weak vector bosons are composite and suggest that larger deviations from
SM values are possible than those suggested here are obliged to provide a
framework for understanding why the SM works at all, including its radiative
corrections.

To sum up, I would reiterate my main conclusions:
\item\bullet  Deviations of anomalous gauge boson couplings from the SM may be
estimated on general grounds and are related to the scale of underlying new
physics.  They are invariably small.
\item\bullet The divergences of loop corrections are not observable and so are
without phenomenological consequences.  There are fine-tuning questions, but
the only naturalness problem arising from loop divergences is, other than the
cosmological constant, the well-known quadratic divergences of scalar particle
masses in nonsupersymmetric theories.

To the extent that the preceding considerations are typical, our best hope to
discover new physics is probably to detect it directly through the production
of new particles or resonances and not, given the anticipated accuracy of
future colliders, via virtual effects.  The possibilities are better, and the
constraints more significant and more easily interpreted, for those higher
order effective interactions (such as four-fermion interactions) that can
arise in tree approximation in an underlying theory, but, unfortunately, the
anomalous triple vector boson couplings are not among them.  As for the
four-vector-boson interactions, their leading behavior is determined solely
by global symmetries and the association of longitudinal vector bosons with
Goldstone bosons.  Since these are derivatively-coupled, their description
involves some scale for the momentum variations which is, if experience in
hadron physics is any guide at all, characteristically also of the order of
1~\tev\ or more.

\medskip
\centerline{\bf{ACKNOWLEDGEMENTS}}

I have enjoyed collaborations with C. Arzt and J. Wudka concerning much
of the work discussed herein.

\refout
\end
\bye